\documentclass{PoS}

\usepackage{bbm}
\usepackage{bm}
\usepackage{graphicx}
\usepackage{epsfig}
\usepackage{subfig}
\usepackage{latexsym}
\usepackage{amsmath}
\usepackage{amsfonts}
\usepackage{amssymb}

\newcommand{\ev}[1]{ \left \langle #1 \right \rangle}
\newcommand{\csw}{c_{\rm{sw}}}

\newcommand{\be}{\begin{equation}}
\newcommand{\ee}{\end{equation}}
\newcommand{\bea}{\begin{eqnarray}} % only untightened
\newcommand{\eea}{\end{eqnarray}}

\title{Mapping the Conformal Window: SU(2) with 4, 6 and 10 flavors of fermions}

\ShortTitle{Mapping the Conformal Window: SU(2) with 4, 6 and 10 flavors of fermions}

\author{Tuomas Karavirta\\
        University of Jyv\"askyl\"a and Helsinki Institute of Physics \\
        E-mail: \email{kari.rummukainen@helsinki.fi}}
        
\author{Kimmo Tuominen\\
        University of Jyv\"askyl\"a and Helsinki Institute of Physics \\
        E-mail: \email{kari.rummukainen@helsinki.fi}}

\author{\speaker{Jarno Rantaharju}\\
        University of Helsinki and Helsinki Institute of Physics\\
        E-mail: \email{jarno.rantaharju@helsinki.fi}}

\author{Kari Rummukainen\\
        University of Helsinki and Helsinki Institute of Physics\\
        E-mail: \email{kari.rummukainen@helsinki.fi}}

\abstract{We present studies of the SU(2) gauge theory with 4, 6 and 10 fermion flavors. These models are expected to lie on both sides of the edge of the conformal window, where the theory has an infrared fixed point. We observe that the coupling grows with the length scale at four flavors, implying QCD-like behavior. At ten flavors the results are compatible with a Bank-Zaks type fixed point. The results at six flavors remain inconclusive: the running is slow towards the infrared but the range and accuracy of the study are insufficient for determining the existence of a fixed point.}

\FullConference{The 30th International Symposium on Lattice Field Theory\\
		 June 24 - 29,  2012\\
		 Cairns, Australia}

\begin{document}

\section{Introduction}

Within the possible phase space of gauge theories there is a group of models with a non-trivial infrared fixed point. In these conformal models, under renormalization group evolution, the coupling runs to smaller values at small distances, exhibiting asymptotic freedom, but runs to a constant at large distances. They have applications in phenomenological model building, such as for technicolor theories \cite{TC,Hill:2002ap,Sannino:2008ha}, where the Higgs sector is replaced with a strongly interacting sector with chiral symmetry breaking. From purely theoretical point of view, mapping the phase diagram of gauge theories in the number of colors $N$ and fermion flavors $N_f$ is interesting for understanding their nonperturbative dynamics from first principles. Many lattice studies of the conformal window have already appeared in the literature: for example SU(2) with fundamental representation fermions \cite{Bursa:2011ru}, SU(2) with adjoint fermions \cite{Catterall:2007yx,Hietanen:2008mr,DelDebbio:
2008zf,Catterall:2008qk,Hietanen:2009az,DeGrand:2011qd} and SU(3) with fermions in the fundamental  \cite{Appelquist:2007hu,Fodor:2009wk,Deuzeman:2008sc} or in the two-index symmetric \cite{Shamir:2008pb}, i.e. the sextet, representation.

\begin{figure}
\centering
\includegraphics[scale=0.5]{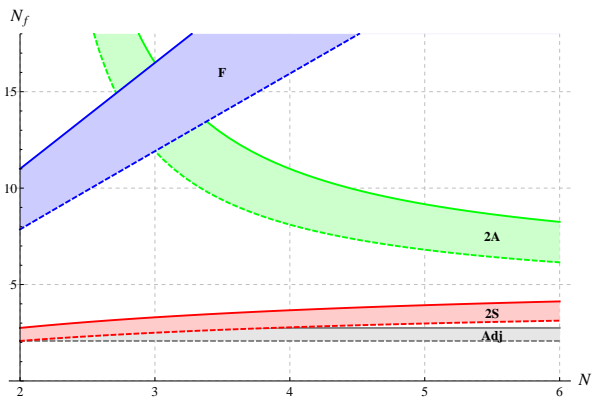}
\caption{The phase diagram of SU($N$) gauge theory as a function of the number of colors, flavors and fermion representations (F = Fundamental,
2A = 2-index antisymmetric, 2S = 2-index symmetric, Adj = Adjoint).
The shaded bands indicate the estimated conformal windows.}
\label{phasediagram}
\end{figure}

In figure \ref{phasediagram} we sketch a phase diagram for SU($N$) gauge theories as a function of $N$ and $N_f$ for model with fermions interacting with the fundamental, two-index (anti)symmetric and adjoint representations of the gauge field. The upper boundary corresponds to the loss of asymptotic freedom, when the first coefficient of the perturbative expansion of the beta function is zero: $\beta_0=11/3 N-4/3 N_f T(R) = 0$, where $T(R)$ is the group theory factor for the fermion representation $R$. Just below the upper bound the value of the fixed point is expected to be small and perturbation theory applicable. When the number of flavors is lowered, however, the fixed point is expected to move to higher coupling. Finally, as one comes to the lower limit, the critical coupling for the chiral symmetry breaking becomes smaller than the expected fixed point and the model becomes chirally broken. 
The lower bound is therefore and inexact approximation in a region where perturbation theory may not be applicable, and needs to be checked using nonperturbative methods. This provides and interesting challenge for the lattice community.

In this study we investigate the phase diagram of SU(2) gauge theory with $N_f  = 10, 6$ and $4$. The results have been published in reference \cite{Karavirta:2011zg}. The models with $10$ and $4$ fermions flavors are expected to lie well within and below the conformal window respectively. The model with $6$ fermions should be close to the lower boundary of the conformal window and is therefore the most challenging of these models.

Since large discretization errors have been observed in previous studies with unimproved Wilson fermions, we measure the coupling using the Schr\"odinger functional method with perturbatively improved Wilson fermions. The models with 4 and 10 fermion flavors behave as expected and are in the confining and conformal phases respectively. Unfortunately, in the 6 flavor case we are unable to resolve whether a fixed point exists, but the possible locations of the fixed point is are at a much higher value of the coupling than suggested by previous unimproved results.

\section{The Method and Results}

The model is defined by the lattice action $S = S_G + S_F$,
where $S_G$ is the standard Wilson plaquette action and $S_F$ is the clover improved fermion action
\begin{equation}
  S_F
  = a^4\sum_{\alpha=1}^{N_f} \sum_x \left [
  \bar{\psi}_\alpha(x) ( i D + m_0 )
  \psi_\alpha(x)
   + a \csw \bar\psi_\alpha(x)\frac{i}{4}\sigma_{\mu\nu}
  F_{\mu\nu}(x)\psi_\alpha(x) \right ],
\end{equation}
where $D$ is the standard Wilson-Dirac operator.
We set the improvement coefficient $\csw$ to the perturbative value \cite{Luscher:1996vw}
$ \csw = 1 + 0.1551(1) g_0^2 + O(g_0^4)$.
We have performed short measurements with $N_f=6$ and $10$ that suggest that this is close to the nonperturbative value at large coupling. This is not the case with $N_f=2$ \cite{Karavirta:2011mv}, where $\csw$ seems to diverge when $g^2$ is increased. We have also included perturbative improvement at the Schr\"odinger functional boundaries,
\begin{align*}
\delta S_{c_t}=& \frac{\beta_L}{4}\sum_{p_T} (c_t -1){\textrm{tr}}(1-U(p)) \\
\delta S_{\tilde{c}_t}=&a^4\sum_x(\tilde
  c_t-1)\frac{1}{a}\bar\psi(x)\psi(x)(\delta(x_0-a)+\delta(x_0-(L-a))
\end{align*}
as described in \cite{Karavirta:2011mv}.

We measure the running coupling using the Schr\"odinger functional method \cite{Luscher:1992an,Luscher:1992ny,Luscher:1993gh,DellaMorte:2004bc}. 
We consider a lattice of volume $V=(Na)^4$. The spatial links at the timelike boundaries of the lattice are fixed to the values
\bea
U_\mu(\bar x,t=0) &= e^{-i\eta \sigma_3 a/L}, \;\;
U_\mu(\bar x,t=L) &= e^{-i(\pi - \eta) \sigma_3 a/L}
\eea
where $\sigma_3$ is the third Pauli matrix.
The spatial boundary conditions are periodic for the gauge field. The fermion fields are set to vanish at the the $t=0$ and $t=L$ boundaries and twisted periodic boundary conditions $\psi(x + L\hat i) = \exp(i\pi/5)\psi(x)$ are set at the spatial boundaries. At the classical level the boundaries generate a constant chromoelectric field and the response of the field to the boundaries can be easily calculated,
\begin{align}
 \frac{\partial S^{\textrm{cl.}}}{\partial \eta} = \frac{k}{g^2_0},
\end{align}
where the constant $k$ is a function of $N=L/a$ and $\eta$ \cite{Luscher:1992ny}.
At full quantum level we define the running coupling $g^2$ trough
\begin{align*}
  \ev{ \frac{\partial S}{\partial \eta} } = \frac{k}{g^2}.
\end{align*}

\begin{figure}
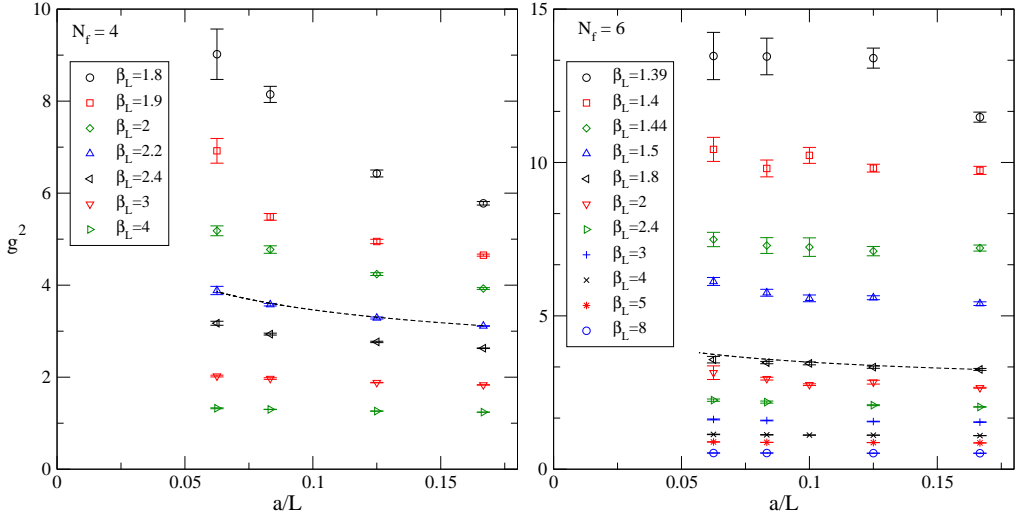

\centering
\includegraphics[height=0.45\textwidth]{nf4g2.eps}
\includegraphics[height=0.45\textwidth]{nf6g2.eps}
\caption[a]{
  The measured values of $g^2(g_0^2,L/a)$ 
  against the inverse lattice size $a/L$ with 4 and 6 flavors of fermions.
  The Black dashed lines give an example of the running in perturbation theory to 2-loop order at a modest coupling, normalized to match the measurement at $L/a=6$.
}
\label{fig:couplingnf46}
\end{figure}

\begin{figure}
\centering
\includegraphics[height=0.45\textwidth]{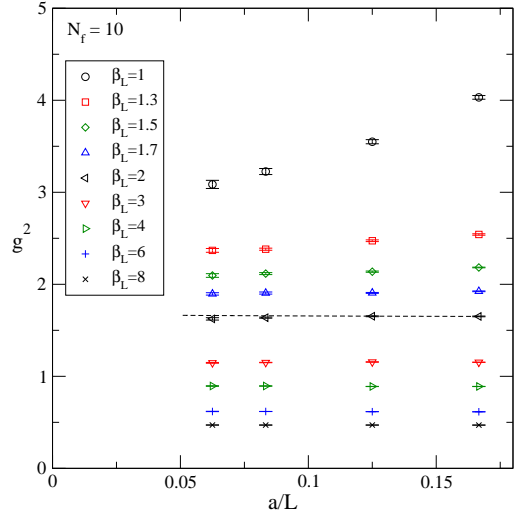}
\caption[a]{
  The measured values of $g^2(g_0^2,L/a)$ 
  against the inverse lattice size $a/L$ with 10 flavors of fermions.
  The Black dashed lines give an example of the running in perturbation theory to 2-loop order at a modest coupling, normalized to match the measurement at $L/a=6$.
}
\label{fig:couplingnf10}
\end{figure}

The measured values of $g^2$ for the models with $N_f=4$ and $6$ are given in figure \ref{fig:couplingnf46} and for the model with $N_f=10$ in figure \ref{fig:couplingnf10}.
The figures show that in the four fermion model the running of the coupling with the energy scale stays negative and increases in magnitude with the coupling. 
In the 10 fermion model the running is slow at small coupling and changes sign between $g^2=1$ or $g^2=2$. 
In the six fermion model the running remains slow to very high coupling but does not seem to change sign, although the within the errors it is impossible to make any conclusion above $g^2=10$.


To quantify the running and facilitate taking the continuum limit, we use the step scaling function $\Sigma(u,s,L/a)$ introduced in \cite{Luscher:1992an}:
\begin{align}
 &\Sigma(u,s,L/a) = \left. g^2(g_0^2,sL/a) \right |_{g^2(g_0^2,L/a)=u}
 \label{stepscaling}\\
 &\sigma(u,s) = \lim_{a/L\rightarrow 0} \Sigma(u,s,L/a).
\end{align}
We choose $s=2$ and calculate $\Sigma(u,s,L/a)$ at $L=6$ and $8$. Since we expect most $\mathcal O(a)$ effects to be absent in the improved model, we obtain a continuum limit using quadratic extrapolation.

For the continuum extrapolation we need to calculate $\Sigma(u,s,L/a)$ at the same measured coupling $u=g^2$ on both lattice sizes. 
We use an interpolating function to define the measured coupling $g^2(g_0^2,L/a)$ in a continuous range of $g_0^2$. At each volume $L/a$ we fit the data to the function
\begin{align*}
  \frac{1}{g^2(g_0^2,L/a)} = \frac{1}{g_0^2} \left[\frac{ 1 + \sum_{i=1}^n
  a_i g_0^{2i}}{1 + \sum_{i=1}^m
  b_i g_0^{2i}}\right].
\end{align*}
For the models with $4$ and $6$ fermion flavors we find the best fit using the parameters $m=2,n=2$ and for the model with 10 flavors $m=1,n=2$.

\begin{figure}
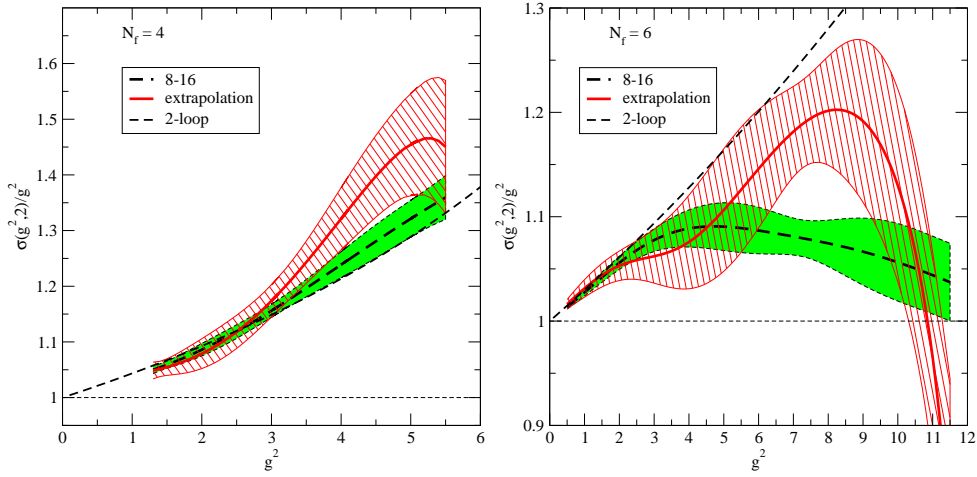

\centering
\includegraphics[scale=0.45]{nf4.eps}
\includegraphics[scale=0.45]{nf6.eps}
\caption[The step scaling function with 4 fermions.]{
  (color online) The scaled step scaling function $\sigma(g^2,2)/g^2$ with 4 and 6 fermions.
  The thick red line corresponds to the continuum extrapolation,
  and the hashed band to the statistical errors of the
  extrapolation. The thick dashed line with the shaded error band is the
  largest volume step scaling function without extrapolation. The thin
  dashed line is the 2-loop perturbative value of $\sigma(g^2,2)/g^2$.
}
\label{sigmanf46}
\end{figure}

\begin{figure}
\centering
\includegraphics[scale=0.4]{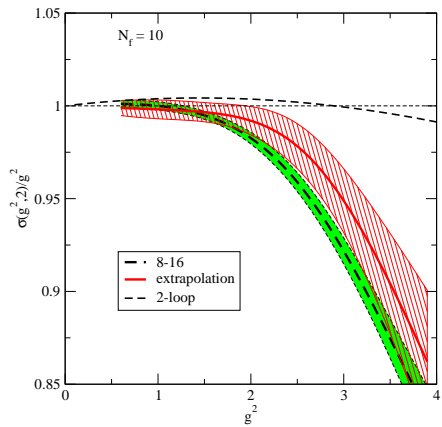}
\caption[The step scaling function with 10 fermions.]{
  As in figure \ref{sigmanf46} but with 10 fermion flavors.}
\label{sigmanf10}
\end{figure}

In figures \ref{sigmanf46} and \ref{sigmanf10} we show the step scaling function for SU(2) with 4, 6 and 10 fermion flavors. Both the models with 4 and 10 fermions behave as expected, with the renormalized step scaling function $\sigma(g^2)/g^2$ increasing with the coupling in the case with 4 fermions and clearly crossing $\sigma(g^2)/g^2=1$ in the case with 10 fermions. In both models the continuum extrapolation starts to deviate from the lattice result at large coupling, implying the presence of discretization effects. The results for the model with 6 fermions are inconclusive: the running remains slow but the discretization errors start to dominate before there is possibility of a fixed point.

To solve the problem we have investigated using smeared actions in combination with $\mathcal O(a)$ improvement to reduce the systematic errors and allow simulations with larger lattice sizes. A hypercubic smearing procedure was used to study the running coupling in \cite{DeGrand:2011qd} and  \cite{DeGrand:2012yq}. It was demonstrated that smearing can reduce the systematic errors and stabilize the simulation.
We are currently investigating SU(2) with two adjoint fermions using a hypercubic stout smearing similar to the one used in \cite{Durr:2010aw,Hasenfratz:2007rf}, using a smeared fermion action and a partially smeared gauge action.

\acknowledgments
J.R. is supported by the Finnish Academy of Science and Letters V\"ais\"al\"a fund. T.K. is supported by the Magnus Ehrnrooth foundation and by University of Jyv\"askyl\"a Faculty of Mathematics and Science. We acknowledge the support from the Academy of Finland grant
number 1134018.  The computations have been performed at the Finnish IT Center for Science.


\begin{thebibliography}{99}

\bibitem{TC} 
S.~Weinberg,
{\em{Implications Of Dynamical Symmetry Breaking: An Addendum}},
Phys.\ Rev.\ D {\bf 19}, 1277 (1979);
%%CITATION = PHRVA,D19,1277;%%
L.~Susskind,
{\em{Dynamics Of Spontaneous Symmetry Breaking In The Weinberg-Salam Theory}},
Phys.\ Rev.\ D {\bf 20}, 2619 (1979).
%%CITATION = PHRVA,D20,2619;%%

  \bibitem{Hill:2002ap}
  C.~T.~Hill and E.~H.~Simmons,
  %``Strong dynamics and electroweak symmetry breaking,''
  Phys.\ Rept.\  {\bf 381}, 235 (2003)
  [Erratum-ibid.\  {\bf 390}, 553 (2004)]
  [arXiv:hep-ph/0203079].
  %%CITATION = PRPLC,381,235;%%

\bibitem{Sannino:2008ha}
  F.~Sannino,
  %``Dynamical Stabilization of the Fermi Scale: Phase Diagram of Strongly
  %Coupled Theories for (Minimal) Walking Technicolor and Unparticles,''
  arXiv:0804.0182 [hep-ph].
  %%CITATION = ARXIV:0804.0182;%%
  
%\cite{Bursa:2011ru}
\bibitem{Bursa:2011ru} 
  F.~Bursa, L.~Del Debbio, D.~Henty, E.~Kerrane, B.~Lucini, A.~Patella, C.~Pica and T.~Pickup {\it et al.},
  %``Improved Lattice Spectroscopy of Minimal Walking Technicolor,''
  Phys.\ Rev.\ D {\bf 84}, 034506 (2011)
  [arXiv:1104.4301 [hep-lat]].
  %%CITATION = ARXIV:1104.4301;%%

\bibitem{Catterall:2007yx}
  S.~Catterall and F.~Sannino,
  %``Minimal walking on the lattice,''
  Phys.\ Rev.\  D {\bf 76}, 034504 (2007)
  [arXiv:0705.1664 [hep-lat]].
  %%CITATION = PHRVA,D76,034504;%%

\bibitem{Hietanen:2008mr}
  A.~J.~Hietanen, J.~Rantaharju, K.~Rummukainen and K.~Tuominen,
  %``Spectrum of SU(2) lattice gauge theory with two adjoint Dirac flavours,''
  JHEP {\bf 0905}, 025 (2009)
  [arXiv:0812.1467 [hep-lat]].
  %%CITATION = JHEPA,0905,025;%%

\bibitem{DelDebbio:2008zf}
  L.~Del Debbio, A.~Patella and C.~Pica,
  %``Higher representations on the lattice: Numerical simulations. SU(2) with
  %adjoint fermions,''
  Phys.\ Rev.\  D {\bf 81}, 094503 (2010)
  [arXiv:0805.2058 [hep-lat]].
  %%CITATION = PHRVA,D81,094503;%%

\bibitem{Catterall:2008qk}
  S.~Catterall, J.~Giedt, F.~Sannino and J.~Schneible,
  %``Phase diagram of SU(2) with 2 flavors of dynamical adjoint quarks,''
  JHEP {\bf 0811}, 009 (2008)
  [arXiv:0807.0792 [hep-lat]].
  %%CITATION = JHEPA,0811,009;%%

\bibitem{Hietanen:2009az}
  A.~J.~Hietanen, K.~Rummukainen and K.~Tuominen,
  %``Evolution of the coupling constant in SU(2) lattice gauge theory with two
  %adjoint fermions,''
  Phys.\ Rev.\  D {\bf 80}, 094504 (2009)
  [arXiv:0904.0864 [hep-lat]].
  %%CITATION = PHRVA,D80,094504;%%

\bibitem{DeGrand:2011qd}
  T.~DeGrand, Y.~Shamir and B.~Svetitsky,
  %``Infrared fixed point in SU(2) gauge theory with adjoint fermions,''
  Phys.\ Rev.\  D {\bf 83}, 074507 (2011)
  [arXiv:1102.2843 [hep-lat]].
  %%CITATION = PHRVA,D83,074507;%%

\bibitem{Appelquist:2007hu}
  T.~Appelquist, G.~T.~Fleming and E.~T.~Neil,
  %``Lattice study of the conformal window in QCD-like theories,''
  Phys.\ Rev.\ Lett.\  {\bf 100}, 171607 (2008)
  [Erratum-ibid.\  {\bf 102}, 149902 (2009)]
  [arXiv:0712.0609 [hep-ph]].
  %%CITATION = PRLTA,100,171607;%%

\bibitem{Fodor:2009wk}
  Z.~Fodor, K.~Holland, J.~Kuti, D.~Nogradi and C.~Schroeder,
  %``Nearly conformal gauge theories in finite volume,''
  Phys.\ Lett.\  B {\bf 681}, 353 (2009)
  [arXiv:0907.4562 [hep-lat]].
  %%CITATION = PHLTA,B681,353;%%

\bibitem{Deuzeman:2008sc}
  A.~Deuzeman, M.~P.~Lombardo and E.~Pallante,
  %``The Physics of eight flavours,''
  Phys.\ Lett.\  B {\bf 670}, 41 (2008)
  [arXiv:0804.2905 [hep-lat]].
  %%CITATION = PHLTA,B670,41;%%

\bibitem{Shamir:2008pb}
  Y.~Shamir, B.~Svetitsky and T.~DeGrand,
  %``Zero of the discrete beta function in SU(3) lattice gauge theory with color
  %sextet fermions,''
  Phys.\ Rev.\  D {\bf 78}, 031502 (2008)
  [arXiv:0803.1707 [hep-lat]].
  %%CITATION = PHRVA,D78,031502;%%

  %\cite{Karavirta:2011zg}
\bibitem{Karavirta:2011zg} 
  T.~Karavirta, J.~Rantaharju, K.~Rummukainen and K.~Tuominen,
  %``Determining the conformal window: SU(2) gauge theory with N_f = 4, 6 and 10 fermion flavours,''
  JHEP {\bf 1205}, 003 (2012)
  [arXiv:1111.4104 [hep-lat]].
  %%CITATION = ARXIV:1111.4104;%%
  
\bibitem{Appelquist:1986an}
  T.~W.~Appelquist, D.~Karabali and L.~C.~R.~Wijewardhana,
  %``Chiral Hierarchies and the Flavor Changing Neutral Current Problem in
  %Technicolor,''
  Phys.\ Rev.\ Lett.\  {\bf 57}, 957 (1986).
  %%CITATION = PRLTA,57,957;%%


\bibitem{Luscher:1996vw}
  M.~Luscher and P.~Weisz,
  %``O(a) improvement of the axial current in lattice QCD to one loop order of
  %perturbation theory,''
  Nucl.\ Phys.\  B {\bf 479}, 429 (1996)
  [arXiv:hep-lat/9606016].
  %%CITATION = NUPHA,B479,429;%%


\bibitem{Karavirta:2011mv}
  T.~Karavirta, A.~Mykkanen, J.~Rantaharju, K.~Rummukainen and K.~Tuominen,
  %``Nonperturbative improvement of SU(2) lattice gauge theory with adjoint or
  %fundamental flavors,''
  JHEP {\bf 1106}, 061 (2011)
  [arXiv:1101.0154 [hep-lat]].
  %%CITATION = JHEPA,1106,061;%%


\bibitem{Luscher:1992an}
  M.~Luscher, R.~Narayanan, P.~Weisz and U.~Wolff,
  %``The Schrodinger functional: A Renormalizable probe for nonAbelian gauge
  %theories,''
  Nucl.\ Phys.\  B {\bf 384}, 168 (1992)
  [arXiv:hep-lat/9207009].
  %%CITATION = NUPHA,B384,168;%%

\bibitem{Luscher:1992ny}
  M.~Luscher, R.~Narayanan, R.~Sommer, U.~Wolff and P.~Weisz,
  %``Determination of the running coupling in the SU(2) Yang-Mills theory from
  %first principles,''
  Nucl.\ Phys.\ Proc.\ Suppl.\  {\bf 30}, 139 (1993).
  %%CITATION = NUPHZ,30,139;%%

\bibitem{Luscher:1993gh}
  M.~Luscher, R.~Sommer, P.~Weisz and U.~Wolff,
  %``A Precise determination of the running coupling in the SU(3) Yang-Mills
  %theory,''
  Nucl.\ Phys.\  B {\bf 413}, 481 (1994)
  [arXiv:hep-lat/9309005].
  %%CITATION = NUPHA,B413,481;%%

%\cite{DellaMorte:2004bc}
\bibitem{DellaMorte:2004bc} 
  M.~Della Morte {\it et al.}  [ALPHA Collaboration],
  %``Computation of the strong coupling in QCD with two dynamical flavors,''
  Nucl.\ Phys.\ B {\bf 713}, 378 (2005)
  [hep-lat/0411025].
  %%CITATION = HEP-LAT/0411025;%%

%\cite{DeGrand:2012yq}
\bibitem{DeGrand:2012yq} 
  T.~DeGrand, Y.~Shamir and B.~Svetitsky,
  %``Mass anomalous dimension in sextet QCD,''
  arXiv:1201.0935 [hep-lat].
  %%CITATION = ARXIV:1201.0935;%%

\bibitem{Durr:2010aw}
  S.~Durr {\it et al.},
  %``Lattice QCD at the physical point: Simulation and analysis details,''
  JHEP {\bf 1108}, 148 (2011)
  [arXiv:1011.2711 [hep-lat]].
  %%CITATION = JHEPA,1108,148;%%

\bibitem{Hasenfratz:2007rf}
  A.~Hasenfratz, R.~Hoffmann and S.~Schaefer,
  %``Hypercubic smeared links for dynamical fermions,''
  JHEP {\bf 0705}, 029 (2007)
  [arXiv:hep-lat/0702028].
  %%CITATION = JHEPA,0705,029;%%
\end{thebibliography}
\end{document}